\documentclass[graybox]{svmult}

\usepackage{mathptmx}       
\usepackage{helvet}         
\usepackage{courier}        
\usepackage{type1cm}        
\usepackage{makeidx}         
\usepackage{graphicx}        
\usepackage{multicol}        
\usepackage[bottom]{footmisc}

\makeindex             

\begin{document}

\title*{The WiFeS S7 AGN survey: Current status and recent results on NGC~6300}
\author{J. Scharw\"achter, M. A. Dopita, P. Shastri, R. Davies, L. Kewley, E. Hampton, R. Sutherland, P. Kharb, J. Jose, H. Bhatt, S. Ramya, 
C. Jin, J. Banfield, I. Zaw, S. Juneau, B. James and S. Srivastava}
\authorrunning{J. Scharw\"achter, M. A. Dopita, P. Shastri, et al.} 
\institute{J. Scharw\"achter \at LERMA, Observatoire de Paris, PSL, CNRS, Sorbonne Universit\'es, UPMC, F-75014 Paris, France \email{julia.scharwaechter@obspm.fr}
\and M. A. Dopita, R. Davies, L. Kewley, E. Hampton, R. Sutherland \at Research School of Astronomy \& Astrophysics, The Australian National University, Cotter Road, Weston Creek, ACT 2611, Australia 
\and P. Shastri, P. Kharb, J. Jose, H. Bhatt, S. Ramya \at Indian Institute of Astrophysics, Koramangala 2B Block, Bangalore 560034, India
\and  C. Jin \at Qian Xuesen Laboratory for Space Technology, Beijing, China
\and J. Banfield \at CSIRO Astronomy \& Space Science, P.O. Box 76, Epping, NSW 1710, Australia
\and I. Zaw \at New York University (Abu Dhabi), 70 Washington Sq. S, New York, NY 10012, USA
\and S. Juneau \at CEA-Saclay, DSM/IRFU/SAp, 91191 Gif-sur-Yvette, France
\and B. James \at Institute of Astronomy, Cambridge University, Madingley Road, Cambridge CB3 0HA, UK
\and S. Srivastava \at Astronomy and Astrophysics Division, Physical Research Laboratory, Ahmedabad 380009, India
}
%
%
\maketitle

\abstract*{The Siding Spring Southern Seyfert Spectroscopic Snapshot Survey (S7) is a targeted survey probing the narrow-line regions (NLRs) of a representative 
sample of $\sim 140$ nearby ($z<0.02$) Seyfert galaxies by means of optical integral field spectroscopy. The survey is based on a homogeneous data set observed using the
Wide Field Spectrograph WiFeS. The data provide a $25\times 38\ \mathrm{arcsec^2}$ field-of-view around the galaxy centre at typically $\sim 1.5$~arcsec spatial resolution and cover
a wavelength range between $\sim 3400 - 7100$~\AA\ at spectral resolutions of $\sim 100\ \mathrm{km\ s^{-1}}$ and $\sim 50\ \mathrm{km\ s^{-1}}$ in the blue and red parts, respectively. The survey is primarily designed to study gas excitation and star formation around AGN, with a special focus on the shape of the AGN ionising continuum, the interaction between radio jets and the NLR gas, and the nature of nuclear LINER emission. We provide an overview of the current status
of S7-based results and present new results for NGC~6300.}

\abstract{The Siding Spring Southern Seyfert Spectroscopic Snapshot Survey (S7) is a targeted survey probing the narrow-line regions (NLRs) of a representative 
sample of $\sim 140$ nearby ($z<0.02$) Seyfert galaxies by means of optical integral field spectroscopy. The survey is based on a homogeneous data set observed using the
Wide Field Spectrograph WiFeS. The data provide a $25\times 38\ \mathrm{arcsec^2}$ field-of-view around the galaxy centre at typically $\sim 1.5$~arcsec spatial resolution and cover
a wavelength range between $\sim 3400 - 7100$~\AA\ at spectral resolutions of $\sim 100\ \mathrm{km\ s^{-1}}$ and $\sim 50\ \mathrm{km\ s^{-1}}$ in the blue and red parts, respectively. The survey is primarily designed to study gas excitation and star formation around AGN, with a special focus on the shape of the AGN ionising continuum, the interaction between radio jets and the NLR gas, and the nature of nuclear LINER emission. We provide an overview of the current status
of S7-based results and present new results for NGC~6300.}

\section{Introduction}
\label{sec:intro}

S7 is a targeted survey based on optical integral field spectroscopy, providing data for 
the central regions of $\sim 140$ nearby Seyfert galaxies on $\sim 100\ \mathrm{pc}$ scales resolution \cite{Dopita:2014aa,Dopita:2014ab,Dopita:2015aa,Dopita:2015ab}.
S7 is designed to probe the AGN NLR by means of emission-line diagnostics and kinematics.
The data set is homogeneous and gives access to a wide wavelength range covering a large number of optical emission lines. 
Line kinematic studies are facilitated by the comparatively high spectral resolution 
of $\sim 50\ \mathrm{km\ s^{-1}}$ in the red part of the spectra.

The S7 sample is a representative sample of nearby Seyfert galaxies, including LINERs and a number of H~{\sc ii} galaxies, with an intentional bias toward radio-detected sources.
Targets are selected from the ``catalogue of quasars and active nuclei'' \cite{Veron-Cetty:2006aa}. 
The sources are typically chosen to have a declination of $\mathrm{DEC}<+10\ \deg$
and to lie outside the Galactic plane by more than $\sim 20\ \deg$. 
In order to guarantee a sufficient physical spatial resolution, the targets are limited to nearby sources ($z<0.02$). For the S7 seeing conditions 
of $\sim 1.5-1.8$~arcsec, this limit implies spatial resolutions of $\sim 100-700$~pc.
Those sources that have available radio data from the NRAO VLA Sky Survey 
(i.e. targets at $\mathrm{DEC}>-41\ \deg$) are typically required to have a minimum flux of $20\ \mathrm{mJy}$ at  1.4~GHz.  

The observations are carried out in a snapshot mode using the Wide Field Spectrograph (WiFeS) operated at the ANU 2.3~m telescope 
at Siding Spring in Australia \cite{Dopita:2007aa}.
WiFeS is an optical integral field spectrograph in image-slicer design with a field-of-view of $25\times 38\ \mathrm{arcsec^2}$. The field-of-view is defined by
25 slitlets of $1\times 38\ \mathrm{arcsec^2}$ size, for which the slit width has been adjusted to match
the typical seeing conditions of $>1\ \mathrm{arcsec}$ at Siding Spring. WiFeS is operated simultaneously in a blue and a red arm so that a wide wavelength coverage is achieved.
For S7, the $R_S=3000$ and $R_S=7000$ gratings are used in the blue and red arm, respectively, which results in a continuous wavelength coverage
between $\sim 3400 - 7100\ \mathrm{\AA}$. The data are reduced using standard procedures provided by the python-based WiFeS reduction pipeline
PyWiFeS \cite{Childress:2014aa}. The reduction process yields calibrated sky-subtracted data cubes sampled on a $1\times1\ \mathrm{arcsec^2}$ pixel scale.
For further details on the observations and data reduction for the first published data set of 64 galaxies, the reader is referred to the main survey paper \cite{Dopita:2015aa}.
Interferometric follow-up observations of S7 sources at radio frequencies using the Australia Telescope Compact Array (ATCA) and the Giant Metrewave Radio 
Telescope (GMRT) in India
are in progress.

\section{Overview of S7 results}

S7 is primarily designed to (i) study the AGN NLR and the mixing between photoionisation from the AGN and from star formation, (ii)
analyse the spectral energy distribution 
of the AGN ionising continuum via detailed photoionisation models for individual sources, and (iii) derive constraints on the nature of the nuclear LINER phenomenon.
Furthermore, the bias toward radio-detected sources has been introduced to add a special focus on the interaction between radio jets and the NLR gas.
In this section, we provide a brief overview over the main science results from the S7 survey so far. These include the first sample paper on 64 S7 sources 
\cite{Dopita:2015aa} and two case studies of S7 galaxies \cite{Dopita:2014aa,Dopita:2015ab}, which showcase the above-mentioned science cases.

Three-colour [O~{\sc iii}]-H$\alpha$-[N~{\sc ii}] images and nuclear spectral properties 
for the first 64 S7 sources are compiled in \cite{Dopita:2015aa}. This information can be used as a reference for source selection from S7 galaxies.
A basic assessment of the occurrence of star forming rings and the relative orientation of ionisation cones is presented as well as a preliminary discussion of
the nature of coronal emission lines. 

The detailed analysis of the Seyfert~2 galaxy NGC~5427, discussed in \cite{Dopita:2014aa}, 
shows an example of S7-based photoionisation models for the NLR. Such models
depend on the gas chemical abundance, the ionisation parameter, and the shape of the AGN ionising continuum.
In \cite{Dopita:2014aa}, the NLR abundance is inferred from the 
chemical abundance measured for H~{\sc ii} regions, resulting in improved constraints on the NLR model.
The NLR model for NGC~5427 is used to derive limits for
the shape of the AGN ionising continuum and to analyse the mixing between
extended NLR emission and background H~{\sc ii} regions.

The nature of nuclear LINER emission is addressed in a case study of the LINER NGC~1052 \cite{Dopita:2015ab}. 
The optical-to-UV emission lines 
in NGC~1052 are interpreted in a double-shock scenario, in which gas already shocked in an accretion flow is subsequently affected
by a cocoon shock driven by the radio jet. This scenario is supported by the WiFeS data for 
the stellar and H$\alpha$ line-of-sight velocities and velocity dispersions in the inner $20\times 20$~arcsec$^{2}$, suggesting
two gas bubbles along the minor axis of the galaxy which are separated by a region of broad line widths.

\section{Extended NLR of NGC~6300}

NGC~6300 ($z=0.0037$, \cite{Mathewson:1996aa}) is a ringed barred spiral galaxy hosting a Seyfert 2 nucleus \cite{Buta:1987aa,Ryder:1996aa,Buta:2001aa}.
The nuclear radio structure at 8.6~GHz has been reported to be slightly resolved, extending over 3.5~arcsec \cite{Morganti:1999aa}. 
Fabry-Perot observations of the large-scale H$\alpha$ velocity field show a blueshift of the nuclear H$\alpha$ emission of
$100\ \mathrm{km\ s^{-1}}$ with respect to the systemic velocity \cite{Buta:2001aa}.
Recent high-spatial-resolution data for the H$_2$ 1-0 S(1) line in the central 
$\sim 4\times 4\ \mathrm{arcsec^2}$ ($\sim 320\times 320\ \mathrm{pc^2}$) show evidence of 
a nuclear biconical outflow in approximately north-south direction with blueshifted velocities to the north and redshifted
velocities to the south \cite{Davies:2014aa}. 

\begin{figure}[b]
\sidecaption
\includegraphics[scale=.65]{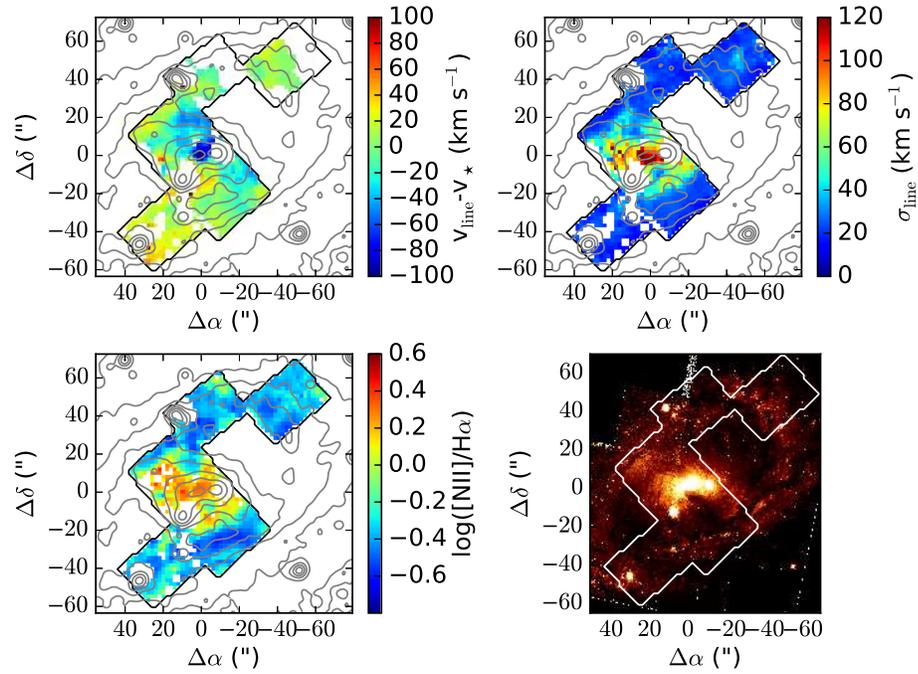}
%
%
\caption{Kinematic and line ratio maps for NGC~6300 derived from the Gaussian fits to [N~{\sc ii}]~$\lambda 6548$, H$\alpha$, and [N~{\sc ii}]~$\lambda 6583$. Upper left panel:
Difference between the emission-line and stellar line-of-sight velocities; Upper right panel:
Intrinsic velocity dispersion of the emission lines; Lower left panel: [N~{\sc ii}]$\lambda 6583$/H$\alpha$ ratios; Lower right panel: Overlay of the 
WiFeS field-of-view on a Hubble Space Telescope WFPC2  image in the $F450W$ filter, taken from the Hubble Legacy Archive. A Digitized Sky Survey (DSS) 
image of NGC~6300 is overlaid on the WiFeS maps
as grey contours. The maps have been clipped using a 3$\sigma$-limit for the line fluxes and a minimum line velocity dispersion of $10\ \mathrm{km\ s^{-1}}$. In addition, a number of foreground stars (visible in the DSS contours) have been masked out.}
\label{fig:1}       
\end{figure}

The WiFeS data for NGC~6300 include six fields covering the nucleus, bar, and parts of the ring (Fig.~\ref{fig:1}). 
The data reduction, fitting and subtraction of the stellar continuum, and the assignment of the World Coordinate System (WCS) is based on similar
methods as described in \cite{Dopita:2014aa}.
The resulting six cubes were resampled into a single cube slice by slice based on the WCS solution using the software {\it Swarp} \cite{Bertin:2002aa} 
with a bilinear interpolation to (binned) output pixels of $2\times2\ \mathrm{arcsec^2}$.

In Fig.~\ref{fig:1}, we present 
the first results from an analysis of H$\alpha$ and [N~{\sc ii}]~$\lambda 6548$ and $\lambda 6583$, 
the strongest lines in the WiFeS spectra for the extended circum-nuclear gas in NGC~6300.
The three lines were fitted simultaneously using a superposition of
single Gaussian profiles per line forced to share the same kinematics. 
The gas and stellar line-of-sight velocities in the nucleus differ by about $\sim 100\ \mathrm{km\ s^{-1}}$ 
(upper left panel of Fig.\ref{fig:1}), in agreement with the H$\alpha$ blueshift reported by \cite{Buta:2001aa}. 
Up to distances of $\sim 20$~arcsec ($\sim 1.6~\mathrm{kpc}$) from the nucleus, the lines are found to have broad profiles ($\sim 60-90\ \mathrm{km\ s^{-1}}$, compared to
$\sim 20-30\ \mathrm{km\ s^{-1}}$ in the H~{\sc ii} regions in the ring) and enhanced [N~{\sc ii}]/H$\alpha$
ratios ($\sim 0.2$, compared to log([N~{\sc ii}]/H$\alpha$)$ < 0$ in the H~{\sc ii} regions), (upper right and lower left panels of Fig.\ref{fig:1}). 
The spectra for the lines east of the nucleus, which appear particularly broad in the single-Gaussian fits, show indications of 
double peaks that suggest more complex kinematics.
The region of enhanced [N~{\sc ii}]/H$\alpha$
ratios indicates ionisation by the AGN continuum, shocks, or a mixture of both, on spatially extended scales around the active nucleus. The correlation of the line ratios
with line width suggests that shocks are likely to play an important role in this region. Shocks could be associated with the nuclear outflow reported in \cite{Davies:2014aa}, a radio jet, 
or with bar-driven perturbations.

\begin{acknowledgement}
J. S. acknowledges the European Research Council
for the Advanced Grant Program Number 267399-Momentum.
\end{acknowledgement}

%
%

\begin{thebibliography}{99.}%

\bibitem{Bertin:2002aa} Bertin, E., Mellier, Y., Radovich, M., Missonnier, G., Didelon, P., \& Morin, B.\ 2002, in Astronomical Data Analysis Software and Systems XI, ed. by D. A. Bohlender, D. Durand, \& T. H. Handley, ASP Conf. Ser., 281, 228
\bibitem{Buta:1987aa} Buta, R.\ 1987, ApJS, 64, 383
\bibitem{Buta:2001aa} Buta, R., Ryder, S.~D., Madsen, G.~J., Wesson, K., Crocker, D.~A., \& Combes, F.\ 2001, AJ, 121, 225
\bibitem{Childress:2014aa} Childress, M.~J., Vogt, F.~P.~A., Nielsen, J., \& Sharp, R.~G.\ 2014, Ap\&SS, 349, 617
\bibitem{Davies:2014aa} Davies, R.~I., Maciejewski, W., Hicks, E.~K.~S., et al.\ 2014, ApJ, 792, 101
\bibitem{Dopita:2007aa} Dopita, M., Hart, J., McGregor, P., Oates, P., Bloxham, G., \& Jones, D.\ 2007, Ap\&SS, 310, 255
\bibitem{Dopita:2014aa} Dopita, M.~A., Scharw{\"a}chter, J., Shastri, P., et al.\ 2014, A\&A, 566, A41
\bibitem{Dopita:2014ab} Dopita, M.~A., Shastri, P., Scharw{\"a}chter, J., et al.\ 2015, in Galaxies in 3D across the Universe, ed. by 
B. L. Ziegler, F. Combes, H. Dannerbauer \& M. Verdugo, Proceedings of the International Astronomical Union, IAU Symposium 309, 10, 200
\bibitem{Dopita:2015aa} Dopita, M.~A., Shastri, P., Davies, R., et al.\ 2015, ApJS, 217, 12
\bibitem{Dopita:2015ab} Dopita, M.~A., Ho, I.-T., Dressel, L.~L., et al.\ 2015, ApJ, 801, 42
\bibitem{Mathewson:1996aa} Mathewson, D.~S., \& Ford, V.~L.\ 1996, ApJS, 107, 97 
\bibitem{Morganti:1999aa} Morganti, R., Tsvetanov, Z.~I., Gallimore, J., \& Allen, M.~G.\ 1999, A\&AS, 137, 457
\bibitem{Ryder:1996aa} Ryder, S.~D., Buta, R.~J., Toledo, H., Shukla, H., Staveley-Smith, L., \& Walsh, W.\ 1996, ApJ, 460, 665
\bibitem{Veron-Cetty:2006aa} V{\'e}ron-Cetty, M.-P., \& V{\'e}ron, P.\ 2006, A\&A, 455, 773

\end{thebibliography}
%

\end{document}